\documentclass[preprint,showpacs,preprintnumbers,amsmath,amssymb]{revtex4}

\usepackage{graphicx}
\usepackage{dcolumn}
\usepackage{bm}
\usepackage{amssymb}

\begin{document}
\newcommand{\vecx}{\mbox{\boldmath $x$}}
\newcommand{\vecp}{\mbox{\boldmath $p$}}

\title{Validity of the factorization approximation and 
correlation induced by nonextensivity
in $N$-unit independent systems
}

\author{Hideo Hasegawa}
\altaffiliation{hideohasegawa@goo.jp}
\affiliation{Department of Physics, Tokyo Gakugei University,  
Koganei, Tokyo 184-8501, Japan}%

\date{\today}

\begin{abstract}
We have discussed the validity of the factorization approximation (FA) 
and nonextensivity-induced correlation, by using the multivariate 
$q$-Gaussian probability distribution function (PDF) 
for $N$-unit identical, independent nonextensive systems. 
The Tsallis entropy is shown to be expressed by 
$S_q^{(N)} = S_{q,FA}^{(N)}+ \Delta S_q^{(N)}$
where $q$ denotes the entropic index, $S_{q,FA}^{(N)}$ a contribution 
in the FA, and $\Delta S_q^{(N)}$ a correction term.
It is pointed out that the correction term of $\Delta S_q^{(N)}$
is considerable for large $\vert q-1 \vert$ and/or large $N$
because the multivariate PDF cannot be expressed 
by the factorized form which is assumed in the FA. This implies that 
the pseudoadditivity of the Tsallis entropy,
which is obtained with PDFs in the FA,
does not hold although it is commonly postulated in the literatures.
We have calculated correlations defined by 
$C_m= \langle (\delta x_i \:\delta x_j)^m \rangle_q
-\langle (\delta x_i)^m \rangle_q\: \langle (\delta x_j)^m \rangle_q$ 
for $i \neq j$, where $\delta x_i=x_i -\langle  x_i \rangle_q$
and $\langle \cdot \rangle_q$ stands for $q$-average over the escort PDF.
It has been shown that $C_1$ expresses the intrinsic correlation and 
that $C_m$ with $m \geq 2$ signifies correlation induced by nonextensivity 
whose physical origin is elucidated within the superstatistics.
PDFs calculated for the classical ideal gas and harmonic oscillator
are compared with the $q$-Gaussian PDF.
A discussion on the $q$-product PDF is presented also.
\end{abstract}

        

\maketitle
\newpage

\section{Introduction}

In the last decade, much attention has been paid to the nonextensive
statistics since Tsallis proposed the so-called Tsallis entropy
\cite{Tsallis88,Tsallis98,Tsallis01,Tsallis04}.
The Tsallis entropy for $N$-unit nonextensive systems is defined by 
\begin{eqnarray}
S_q^{(N)} &=& \frac{k_B}{q-1}
\left( 1-\int \left[p_q^{(N)}(\vecx)\right]^q \:d \vecx \right),
\label{eq:A1}
\end{eqnarray}
where $q$ is the entropic index, $k_B$ the Boltzmann constant
($k_B=1$ hereafter), $p_q^{(N)}(\vecx)$ denotes the $N$-variate probability
distribution function (PDF), $\vecx = \{ x_i \}$ ($i=1$ to $N$), and 
$d \vecx = \prod_{i=1}^N dx_i $.
The Tsallis entropy is a one-parameter generalization of
the Boltzmann-Gibbs entropy, to which the Tsallis entropy
reduces in the limit of $q \rightarrow 1.0$.
The Tsallis entropy is nonextensive (non-additive), 
which is shown in the literatures as follows 
\cite{Tsallis88,Tsallis98,Tsallis01,Tsallis04}. 
When the PDF for two independent subsystems $A$ and $B$ is factorized
into those of $A$ and $B$ ($x_1 \in A,\; x_2 \in B$),
\begin{eqnarray}
p_{q}^{(2)}(x_1, x_2) &=& p_q^{(1)}(x_1) \:p_q^{(1)}(x_2),
\label{eq:A11}
\end{eqnarray} 
Eq. (\ref{eq:A1}) yields 
\begin{eqnarray}
S_{q}^{(2)}(A+B) &=& S_q^{(1)}(A)+S_q^{(1)}(B) 
+(1-q) S_q^{(1)}(A) \:S_q^{(1)}(B), 
\label{eq:A2}
\end{eqnarray}
which is referred to as the $pseudoadditive$ relation.
When the PDF for $N$-unit independent subsystems is given as factorized 
form,
\begin{eqnarray}
p_{q}^{(N)}(\vecx) &=& \prod_{i=1}^N \: p_q^{(1)}(x_i),
\label{eq:A4}
\end{eqnarray}
we obtain the pseudoadditive Tsallis entropy $S_q^{(N)}$ 
which is expressed by
\begin{eqnarray}
\ln \left[1+(1-q) S_{q}^{(N)} \right] 
&=& \sum_{i=1}^N \ln \left[1+(1-q) S_{q}^{(1)}(i) \right]. 
\label{eq:D10}
\end{eqnarray}
It should be, however, noted that Eqs. (\ref{eq:A2}) 
and (\ref{eq:D10}) are not correct
in the strict sense because the bivariate PDF derived by the
maximum-entropy method (MEM) cannot be expressed 
by Eq. (\ref{eq:A11}) or (\ref{eq:A4}), as will be shown shortly [Eq. (\ref{eq:C9})].
Indeed, our calculation for identical, independent systems
with the use of exact multivariate PDFs to Eq. (\ref{eq:A1})
yields
\begin{eqnarray}
S_{q}^{(2)} &=& 2 S_q^{(1)}+(1-q)\left[S_q^{(1)}\right]^2
+\Delta S_q^{(2)}, 
\label{eq:A12}\\
S_q^{(N)} &=& S_{q,FA}^{(N)} + \Delta S_q^{(N)},
\label{eq:A3}
\end{eqnarray}
where $S_{q,FA}^{(N)}$ denotes $S_{q}^{(N)}$ in Eq. (\ref{eq:D10})
evaluated by the PDF of Eq. (\ref{eq:A4})
in the factorized approximation (FA),
and $\Delta S_q^{(N)}$ expresses a correction term [Eq. (\ref{eq:D6})].
Equations (\ref{eq:A12}) and (\ref{eq:A3}) show that $S_{q}^{(N)}$
does not satisfy the pseudoadditivity.

The PDF is evaluated by the MEM 
for the Tsallis entropy with imposing some constraints.
At the moment, there are four possible MEMs: 
(a) original method \cite{Tsallis88},
(b) un-normalized method \cite{Curado91}, 
(c) normalized method \cite{Tsallis98}, and 
(d) the optimal Lagrange multiplier (OLM) method \cite{Martinez00}.
A comparison among the four MEMs is made in Ref. \cite{Tsallis04}.
Although the four methods are equivalent in the sense that PDFs 
derived in them are easily transformed to each other \cite{Ferri05},
obtained expressions for physical quantities are
ostensibly different depending on the adopted MEM.

Let us consider $N$-unit independent systems whose hamiltonian is given by
\begin{eqnarray}
H &=& \sum_{i=1}^N \: h_i.
\label{eq:A5}
\end{eqnarray}
PDFs for the hamiltonian in the Boltzmann-Gibbs statistics ($q=1.0$) 
may be calculated with the use of either $H$ or $h_i$ because they are given by
\begin{eqnarray}
p_1^{(N)}(\vecx) &\propto& {\rm Tr} \;e^{-\beta H} 
= \prod_{i=1}^N\; p_1^{(1)}(x_i),  \\
p_1^{(1)}(x_i) &\propto&  {\rm Tr}_i \;e^{\beta h_i},
\end{eqnarray}
where $\beta$ is the inverse of temperature, Tr denotes the full trace 
and ${\rm Tr}_i$ the partial trace over $i$. 
It is, however, not the case in the nonextensive statistics in which PDFs 
are given by
\begin{eqnarray}
p_q^{(N)}(\vecx) &\propto& {\rm Tr} \;\exp_q\left(- \beta H \right), 
\label{eq:A10}\\ 
&\neq& \prod_{i=1}^N \;p_q^{(1)}(x_i) 
\hspace{2cm}\mbox{for $q \neq 1.0$}, 
\label{eq:A6} \\
p_q^{(1)}(x_i)&\propto & {\rm Tr}_i \;\exp_q(-\beta h_i).
\label{eq:A7}
\end{eqnarray}
Here $\exp_q(x)$ expresses the $q$-exponential function defined by
\begin{eqnarray}
\exp_q(x) &=& 
[1+(1-q)x]_{+}^{1/(1-q)},
\label{eq:A8}
\end{eqnarray}
with $[x]_{+} ={\rm max}(x,0)$, which reduces to $\exp_q(x)= e^x$ 
for $q \rightarrow 1.0$.
The inequality in Eq. (\ref{eq:A6}) arises from the properties of
the $q$-exponential function,
\begin{eqnarray}
\exp_q(x+y) &\neq& \exp_q(x) \exp_q(y)
\hspace{2cm}\mbox{for $q \neq 1.0$}.
\label{eq:A9}
\end{eqnarray}
Then it has been controversial whether we should employ $H$ or $h_i$
in calculating the PDF in the nonextensive statistics.
In many applications of the nonextensive statistics,
one usually calculates $p_1(x_i)$ with the use of $h_i$ in Eq. (\ref{eq:A7}), 
explicitly or implicitly employing the FA, 
because an exact evaluation of $p_N(\vecx)$ in Eq. (\ref{eq:A10}) 
is generally difficult. This issue of the degree of freedom $N$ on the PDF 
has been discussed in Refs. \cite{Wang02,Wang02b,Jiulin09}.
A calculation for $N$-unit harmonic oscillator shows that the partition
function obtained in the FA is quite different from that obtained by the
exact PDF \cite{Lenzi01}, related discussion being given in Sec. IIIC.

In our previous papers \cite{Hasegawa08b,Hasegawa09}, we discussed
the effect of spatial correlation on the Tsallis entropy and
the generalized Fisher information in nonextensive systems. 
We obtained the multivariate $q$-Gaussian PDF with the OLM-MEM \cite{Martinez00}, 
which correctly includes correlation. 
It is the purpose of the present paper to discuss
the issue mentioned above, by using the exact multivariate
$q$-Gaussian PDF derived in Ref. \cite{Hasegawa08b}.
One of the advantages of a use of the $q$-Gaussian PDF is 
that it is free from an ambiguity in defining the physical temperature
in conformity with the zeroth law of thermodynamics 
in the nonextensive statistics \cite{Tsallis04}.
From calculations of the Tsallis entropy and correlation
in $N$-unit independent systems,
we will show the importance of effects
which are not taken into account in the FA.
The nonextensivity-induced correlation has been discussed 
for classical ideal gas \cite{Abe99b,Liyan08,Feng10}
and harmonic oscillator \cite{Liyan08}.

The superstatistics is one of alternative approaches to the nonextensive 
statistics besides the MEM \cite{Wilk00,Beck01,Beck05}
(for a recent review, see \cite{Beck07}).
In the superstatistics, it is assumed that {\it locally} the equilibrium state 
of a given system is described by the Boltzmann-Gibbs statistics and
its global properties may be expressed by a superposition over the fluctuating 
intensive parameter ({\it i.e.,} the inverse temperature) 
\cite{Wilk00}-\cite{Beck07}.
The superstatistics has been adopted in many kinds 
of subjects such as hydrodynamic turbulence, 
cosmic ray 
and solar flares \cite{Beck07}. 
The physical origin of the nonextensivity-induced correlation
may be elucidated within the superstatistics.

The paper is organized as follows. In Sec. II, 
multivariate PDFs for correlated nonextensive systems
derived by the OLM-MEM \cite{Martinez00} are briefly discussed
\cite{Hasegawa08b,Hasegawa09}.
By using the multivariate PDF, we calculate the Tsallis entropy and correlations.
Some model calculations of the $q$- and $N$-dependent Tsallis entropy 
and correlations are presented. In Sec. III, we discuss the physical origin 
of the nonextensivity-induced correlation, calculating the PDF  
within the superstatistics \cite{Wilk00,Beck01}.
PDFs of the one-dimensional classical ideal gas and harmonic oscillators
derived with the use of the OLM-MEM  \cite{Martinez00}
are compared with the $q$-Gaussian PDF.
The PDF expressed by the $q$-product \cite{Borges04} is also
discussed. Sec. IV is devoted to our conclusion.

\section{$q$-Gaussian PDF}
\subsection{OLM-MEM}

We consider $N$-unit nonextensive systems whose PDF, $p_q^{(N)}(\vecx)$, 
is derived with the use of the OLM-MEM \cite{Martinez00} for the Tsallis 
entropy given by Eq. (\ref{eq:A1}) \cite{Tsallis88,Tsallis98}.
We impose four constraints given by 
(for details, see Appendix B of Ref. \cite{Hasegawa08b})
\begin{eqnarray}
1 &=& \int p_q^{(N)}(\vecx)\:d \vecx, 
\label{eq:B3}
\\
\mu &=& \frac{1}{N}\sum_{i=1}^N \langle x_i \rangle_q, 
\label{eq:B4}\\
\sigma^2 &=& \frac{1}{N} \sum_{i=1}^N
\langle (x_i-\mu)^2 \rangle_q, \\ 
\label{eq:B5} 
s \:\sigma^2 &=& \frac{1}{N(N-1)} \sum_{i=1}^N \sum_{j=1 (\neq i)}^N
\langle (x_i-\mu)(x_j-\mu) \rangle_q.
\label{eq:B6}
\end{eqnarray}
Here $\mu$, $\sigma^2$ and $s$ express the mean, variance, and degree of 
intrinsic correlation, respectively, and $\langle \cdot \rangle_q$ denotes 
the $q$-average over the escort PDF,
\begin{eqnarray}
P_q^{(N)}(\vecx) &=& \frac{\left[p_q^{(N)}(\vecx) \right]^q} {c_q^{(N)}}, 
\end{eqnarray}
with
\begin{eqnarray}
c_q^{(N)} &=& \int \left[p_q^{(N)}(\vecx) \right]^q \:d \vecx.
\label{eq:B7}
\end{eqnarray}
Evaluations of the $q$-average with the use of the exact approach
\cite{Prato95,Rajagopal98} are discussed in Appendix A.

The OLM-MEM with the constraints given by 
Eqs. (\ref{eq:B3})-(\ref{eq:B6}) leads to the PDF
given by \cite{Hasegawa08b}
\begin{eqnarray}
p_q^{(N)}(\vecx) &=& \frac{1}{Z_q^{(N)}}
\exp_q\left[- \left( \frac{1}{2 \nu_q^{(N)} \sigma^2 }\right)
\Phi(\vecx)  \right],
\label{eq:C1}
\end{eqnarray}
where
\begin{eqnarray}
\Phi(\vecx) &=& \sum_{i=1}^N \sum_{j=1}^N 
[a_0 \: \delta_{ij}+ a_1 (1-\delta_{ij})]
(x_i-\mu)(x_j-\mu), 
\label{eq:C0} \\
a_0 &=& \frac{[1+(N-2)s]}{(1-s)[1+(N-1)s]},\\
a_1 &=& - \frac{s}{(1-s)[1+(N-1)s]},
\end{eqnarray}
\begin{eqnarray}
Z_q^{(N)} = \left\{ \begin{array}{ll}
r_s^{(N)} \left[\frac{2 \pi \nu_q^{(N)} \sigma^2}{q-1} \right]^{N/2}
\frac{\Gamma\left(\frac{1}{q-1}-\frac{N}{2} \right)}
{\Gamma \left(\frac{1}{q-1} \right)}
\quad & \mbox{for $q > 1 $}, \\ 
r_s^{(N)}(2 \pi \sigma^2)^{N/2}
\quad & \mbox{for $q=1$},  \\
%
r_s^{(N)} \left[\frac{2\pi \nu_q^{(N)} \sigma^2}{1-q}\right]^{N/2}
\frac{\Gamma\left(\frac{1}{1-q}+1 \right)}
{\Gamma \left(\frac{1}{1-q}+1+\frac{N}{2} \right)}
\quad & \mbox{for $ q <1$},  
\end{array} \right. 
\label{eq:C2}
\end{eqnarray}
\begin{eqnarray}
r_s^{(N)} &=& \{(1-s)^{N-1}[1+(N-1)s]  \}^{1/2}, 
\label{eq:C4}\\
\nu_q^{(N)} &=& \frac{(N+2)-Nq}{2},
\label{eq:C5}
\end{eqnarray}
$B(x,y)$ and $\Gamma(z)$ denoting the beta and gamma functions, respectively. 
Hereafter we assume that the entropic index $q$ takes a value,
\begin{eqnarray}
0 < q < 1+\frac{2}{N},
\label{eq:C12}
\end{eqnarray}
because $p_q^{(N)}(\vecx)$ given by Eq. (\ref{eq:C1}) has the probability properties
with $\nu_q^{(N)} > 0$ for $q < 1+2/N $ 
and because the Tsallis entropy is stable for $q > 0$ 
\cite{Abe02}.

In the absence of the intrinsic correlation ($s=0$) for which
\begin{eqnarray}
\Phi(\vecx) &=& \sum_i \:(x_i-\mu)^2,
\label{eq:C11}
\end{eqnarray} 
Eq. (\ref{eq:C1}) reduces to
\begin{eqnarray}
p_q^{(N)}(\vecx) &=& \frac{1}{Z_q^{(N)}}
\exp_q\left[- \left( \frac{1}{2 \nu_q^{(N)} \sigma^2 }\right) 
\sum_{i=1}^N \:(x_i-\mu)^2 \right]. 
\label{eq:C6}
\end{eqnarray}
On the other hand, the PDF in the FA is given by 
%
\begin{eqnarray}
p_{q,FA}^{(N)}(\vecx) &=& \prod_{i=1}^N \:p_q^{(1)}(x_i), \\
&=&  \frac{1}{\left( Z_q^{(1)} \right)^N}  \prod_{i=1}^N 
\exp_q \left[-\left(\frac{1}{2 \nu_q^{(1)} \sigma^2}\right) 
\: (x_i-\mu)^2 \right].
\label{eq:C8}
\end{eqnarray}
In Eqs. (\ref{eq:C6}) and (\ref{eq:C8}), $Z_q^{(N)}$ 
is given by Eq. (\ref{eq:C2}) with $r_s^{(N)}=1.0$.
From a comparison between Eqs. (\ref{eq:C6}) and (\ref{eq:C8}), 
it is evident that
\begin{eqnarray}
p_q^{(N)}(\vecx) & \neq & p_{q,FA}^{(N)}(\vecx),
\label{eq:C9}
\end{eqnarray}
except for $q=1.0$ or $N=1$.

\subsection{Tsallis entropy}

Substituting the PDF given by Eqs. (\ref{eq:C1})-(\ref{eq:C5}) 
to Eq. (\ref{eq:A1}), we first calculate the Tsallis entropy 
which is given by \cite{Hasegawa08b}
\begin{eqnarray} 
S_q^{(N)} &=& \frac{1-c_q^{(N)}}{q-1}, 
\label{eq:D1}
\end{eqnarray}
with
\begin{eqnarray}
c_q^{(N)} &=& \nu_q^{(N)} \left[Z_q^{(N)} \right]^{1-q}.
\label{eq:D2}
\end{eqnarray}
The $s$-dependence of the Tsallis entropy was previously discussed 
(see Fig. 1 of Ref. \cite{Hasegawa08b}).
With increasing $s$, the Tsallis entropy is decreased as given by
\begin{eqnarray}
S_q^{(N)}(s) &=& S_q^{(N)}(0)-\frac{N(N-1)c_q^{(N)}}{4}\:s^2
\hspace{1cm}\mbox{for $\vert s \vert \ll 2/\sqrt{N(N-1)}$ }.
\end{eqnarray}

Now we pay our attention to identical, independent systems with $s=0$, 
for which we obtain
\begin{eqnarray} 
S_q^{(N)} &=& S_{q,FA}^{(N)} + \Delta S_q^{(N)},
\label{eq:D3} 
\end{eqnarray}
with
\begin{eqnarray}
S_{q,FA}^{(N)} &=& \frac{1-(c_q^{(1)} )^N}{q-1},
\label{eq:D5} \\
\Delta S_q^{(N)} &=& 
\left( \frac{1}{q-1} \right) [(c_q^{(1)})^N -c_q^{(N)}], \\
&=& \left( \frac{1}{q-1} \right) 
\left( \left[\nu_q^{(1)} (Z_q^{(1)})^{1-q} \right]^N 
-\nu_q^{(N)} (Z_q^{(N)})^{1-q} \right).
\label{eq:D6}
\end{eqnarray}
Here $S_{q,FA}^{(N)}$ denotes the Tsallis entropy calculated in the FA, 
and $\Delta S_q^{(N)}$ signifies a correction term.
Equation (\ref{eq:D10}) leads to
\begin{eqnarray}
S_{q,FA}^{(N)} &=& 
\sum_{k=1}^N \frac{N!}{(N-k)! \:k!}
(1-q)^{k-1} (S_q^{(1)})^k, 
\label{eq:D11}\\
&=& N S_q^{(1)} + \frac{N(N-1)(1-q)}{2} (S_q^{(1)})^2 + \cdot\cdot.
\end{eqnarray}

In particular for $N=2$, Eq. (\ref{eq:D3}) becomes
\begin{eqnarray}
S_q^{(2)} &=& \frac{1-c_q^{(2)}}{q-1} 
= S_{q,FA}^{(2)} + \Delta S_q^{(2)}, 
\label{eq:D7}
\end{eqnarray}
with
\begin{eqnarray}
S_{q,FA}^{(2)} &=& \frac{1-(c_q^{(1)})^2}{q-1}
=2 S_q^{(1)} +(1-q) (S_q^{(1)})^2, 
\label{eq:D8}\\
\Delta S_q^{(2)} &=& 
\left( \frac{1}{q-1} \right) [(c_q^{(1)})^2 -c_q^{(2)}], \\
&=& \left( \frac{1}{q-1} \right) 
\left( \left[\nu_q^{(1)} (Z_q^{(1)})^{1-q} \right]^2 
- \nu_q^{(2)} (Z_q^{(2)})^{1-q} \right).
\label{eq:D9}
\end{eqnarray}

\begin{figure}
\begin{center}
\includegraphics[keepaspectratio=true,width=100mm]{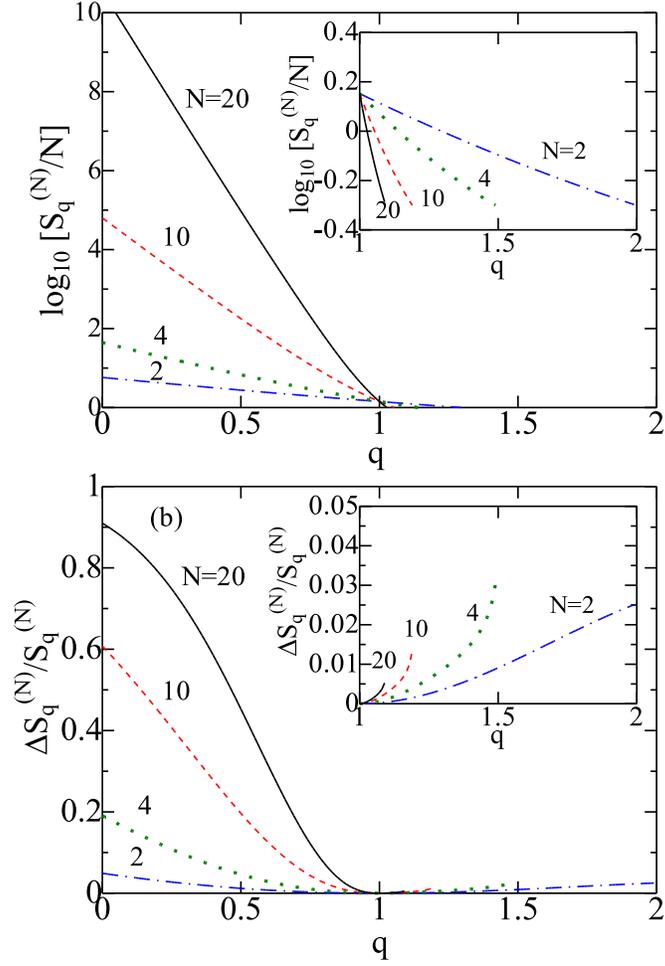}
\end{center}
\caption{
(Color online)
The $q$ dependence of (a) $S_q^{(N)}/N$ and (b) $\Delta S_q^{(N)}/S_q^{(N)}$
for $N=2$ (chain curves), 4 (dotted curves), 10 (dashed curves) 
and 20 (solid curves): insets show enlarged plots
for $q \geq 1.0$: the ordinates in (a) are in the
logarithmic scale.
}
\label{fig1}
\end{figure}

\begin{figure}
\begin{center}
\includegraphics[keepaspectratio=true,width=100mm]{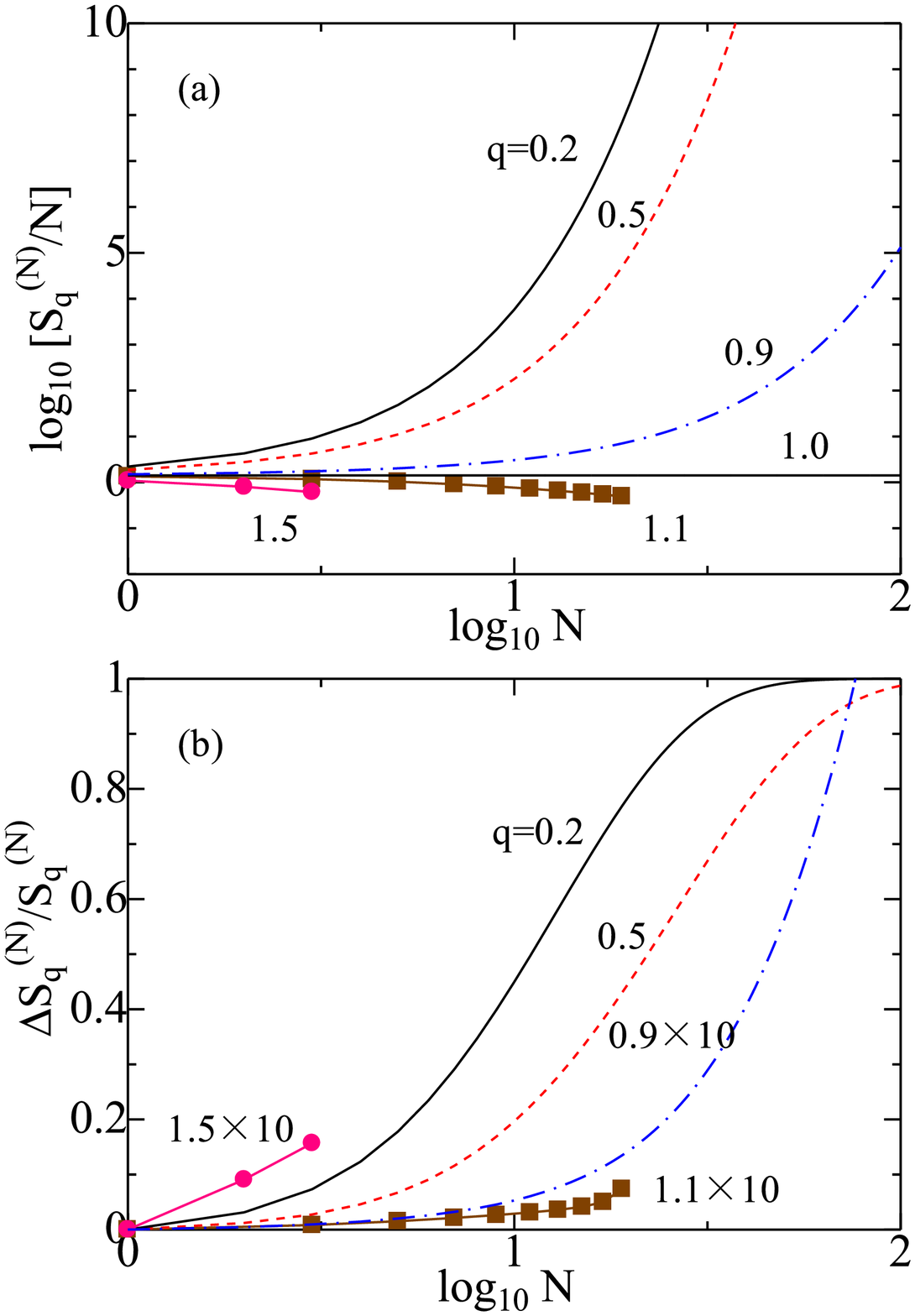}
\end{center}
\caption{
(Color online)
(a) $\log_{10}[S_q^{(N)}/N]$ and (b) $\Delta S_q^{(N)}/S_q^{(N)}$
as a function of $\log_{10} N$ for $q=0.2$ (solid curves), 
0.5 (dashed curves), 0.9 (chain curves), 1.1 (squares) and 1.5 (circles): 
results for $q=0.9$, 1.1 and 1.5 in (b) are multiplied by a factor of ten.
Note that, for $q=1.0$, $S_q^{(N)}/N =1.4189$ and 
$\Delta S_q^{(N)}/S_q^{(N)} =0$ independently of $N$.
}
\label{fig2}
\end{figure}

We will show some model calculations of $S_q^{(N)}$ and $\Delta S_q^{(N)}$.
The $q$ dependences of $\log_{10} (S_q^{(N)}/N)$ for various $N$ values 
are shown in Fig. 1(a) where the inset shows an enlarged plot for $q \geq 1.0$.
With decreasing $q$ from unity, $S_q^{(N)}$ is logarithmically increased. 
The $N$ dependence of $\Delta S_q^{(N)}/S_q^{(N)}$ is shown in
Fig. 1(b) where the inset again shows an enlarged plot for $q \geq 1.0$. 
We note that $\Delta S_q^{(2)}/S_q^{(2)}$ is 0.049, 0.0 and 0.025  
for $q=0.0$, 1.0 and 2.0, respectively.
When $N$ is more increased, $\Delta S_q^{(N)}/S_q^{(N)}$ becomes more considerable.
For example, its value becomes 0.61 and 0.91 for $N=10$ and 20, 
respectively, at $q=0.0$.

The $N$ dependence of $\log_{10} (S_q^{(N)}/N)$ is shown in Fig. 2(a).
We note that with increasing $N$, $S_q^{(N)}$ is increased (decreased)
for $q < 1.0$ ($q > 1.0$).
The $N$ dependence of $\Delta S_q^{(N)}/S_q^{(N)}$ is shown in Fig. 2(b) where
results for $q=0.9$, 1.1 and 1.5 are multiplied by a factor of ten.
It is realized that for $q=0.2$ and 0.5, 
$\Delta S_q^{(N)}/S_q^{(N)} \rightarrow 1.0$ as $N \rightarrow 100$ where
$S_{q,FA}^{(N)}$ almost completely underestimates the Tsallis entropy.

Figures 1(b) and 2(b) clearly show that $\Delta S_q^{(N)}/S_q^{(N)}$ 
is positive and becomes appreciable for large values of $N$ 
and/or large $\vert q-1 \vert $, in particular for $q < 1$.

\subsection{Correlation}

Next we calculate correlations for $i \neq j$ defined by
\begin{eqnarray}
C_{m} &\equiv & \langle (\delta x_i \:\delta x_j)^{m} \rangle_q
-\langle (\delta x_i)^{m} \rangle_q\: \langle (\delta x_i)^{m} \rangle_q, 
\label{eq:E0}
\end{eqnarray}
where $\delta x_i=x_i - \mu$. With the use of the PDF given 
by Eqs. (\ref{eq:C1})-(\ref{eq:C5}), we obtain the first- and second-order
correlations given by (for details, see Appendix A)
\begin{eqnarray}
C_1 
&=& \sigma^2 s, 
\label{eq:E1}\\
C_2 
&=& C_{2s} + C_{2n},
\label{eq:E3}
\end{eqnarray}
where
\begin{eqnarray}
C_{2s} &=& \frac{ 2 [(N+2)-Nq] \:\sigma^4 s^2}
{[(N+4)-(N+2)q ]},
\label{eq:E4}\\
&=& \left\{ \begin{array}{ll}
0
\quad & \mbox{for $s=0$}, \\
\infty
\quad & \mbox{for $q \rightarrow 1+2/(N+2)$}, \\
2 s^2
\quad & \mbox{for $q=1.0$ or $N \rightarrow \infty$},  \\
\end{array} \right. \\
C_{2n} &=& \frac{2(q-1) \:\sigma^4}
{[(N+4)-(N+2)q ]}, 
\label{eq:E5}\\
&=& \left\{ \begin{array}{ll}
0
\quad & \mbox{for $q \rightarrow 1.0$}, \\
\infty
\quad & \mbox{for $q \rightarrow 1+2/(N+2)$}, \\
-\frac{2}{N}
\quad & \mbox{for $N \rightarrow \infty$}. \\
\end{array} \right.
\end{eqnarray} 
We note that $C_1$ and $C_{2s}$ arise from
intrinsic correlation $s$ for $0 \leq q < 1+2/(N+2)$.
In contrast, $C_{2n}$ expresses correlation induced by nonextensivity, 
which vanishes for $q=1.0$ and which approaches
$-2/N$ as $N \rightarrow \infty$.
In particular for $N=2$, $C_{2s}$ and $C_{2n}$ are given by
\begin{eqnarray}
C_{2s} &=& \frac{2 (2-q) \sigma^4 s^2}{(3-2q) }, \\
C_{2n} &=& \frac{(q-1)\:\sigma^4}{(3-2q) }.
\end{eqnarray}
On the contrary, the factorized PDF given by Eq. (\ref{eq:C8}) yields
\begin{eqnarray}
C_1^{FA} &=& C_{2s}^{FA} = C_{2n}^{FA} =0.
\end{eqnarray}

By using Eq. (\ref{eq:C6}), we obtain correlation
of $C_m$ for arbitrary $m$ with $s=0$,
\renewcommand{\arraystretch}{1.5}  
\begin{eqnarray}
C_m 
&=& \left\{ \begin{array}{ll}
A_m \left[\frac{2 \nu_q^{(N)} \sigma^2}{(q-1)} \right]^{m}
\left[\frac{\Gamma(q/(q-1)-N/2-m)}{\Gamma(q/(q-1)-N/2)}
-\left(\frac{\Gamma(q/(q-1)-N/2-m/2)}{\Gamma(q/(q-1)-N/2)} \right)^2 \right]
\quad & \mbox{for $q > 1.0$}, \\
A_m \left[\frac{2 \nu_q^{(N)} \sigma^2}{(1-q)} \right]^{m}
\left[\frac{\Gamma(q/(1-q)+N/2+1)}{\Gamma(q/(1-q)+N/2+m+1)} 
-\left(\frac{\Gamma(q/(1-q)+N/2+1)}{\Gamma(q/(1-q)+N/2+m/2+1)} \right)^2
\right]
\quad & \mbox{for $q <  1.0$}, 
\end{array} \right.  \nonumber \\
&&
\end{eqnarray}
where
\begin{eqnarray}
A_m &=& \left\{ \begin{array}{ll}
\left[\frac{\Gamma(1/2+m/2)}{\Gamma(1/2)} \right]^2
\quad & \mbox{for even $m$}, \\
0
\quad & \mbox{for odd $m$}.
\end{array} \right.
\end{eqnarray}
The nonextensivity yields higher-order correlation
of $C_m$ than $m \geq 2$ for $q \neq 1.0$ 
in independent nonextensive systems where $s=C_1=0$. 

\begin{figure}
\begin{center}
\includegraphics[keepaspectratio=true,width=100mm]{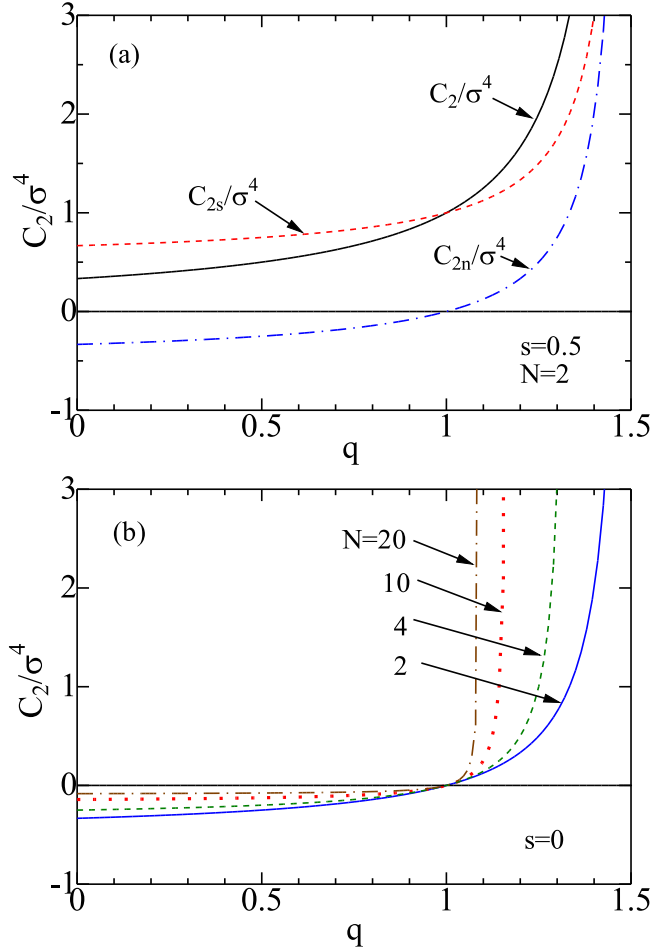}
\end{center}
\caption{
(Color online)
(a) The $q$ dependence of $C_{2s}$ (the dashed curve),
$C_{2n}$ (the dashed curve) and $C_2$ (=$C_{2s}+C_{2n}$, the solid curve)
for $s=0.5$ and $N=2$.
(b) The $q$ dependence of $C_2$ with $s=0$ for $N=2$ (the solid curve),
4 (the dotted curve), 10 (the dotted curve) and 20 (the chain curve).
}
\label{fig3}
\end{figure}

\begin{figure}
\begin{center}
\includegraphics[keepaspectratio=true,width=100mm]{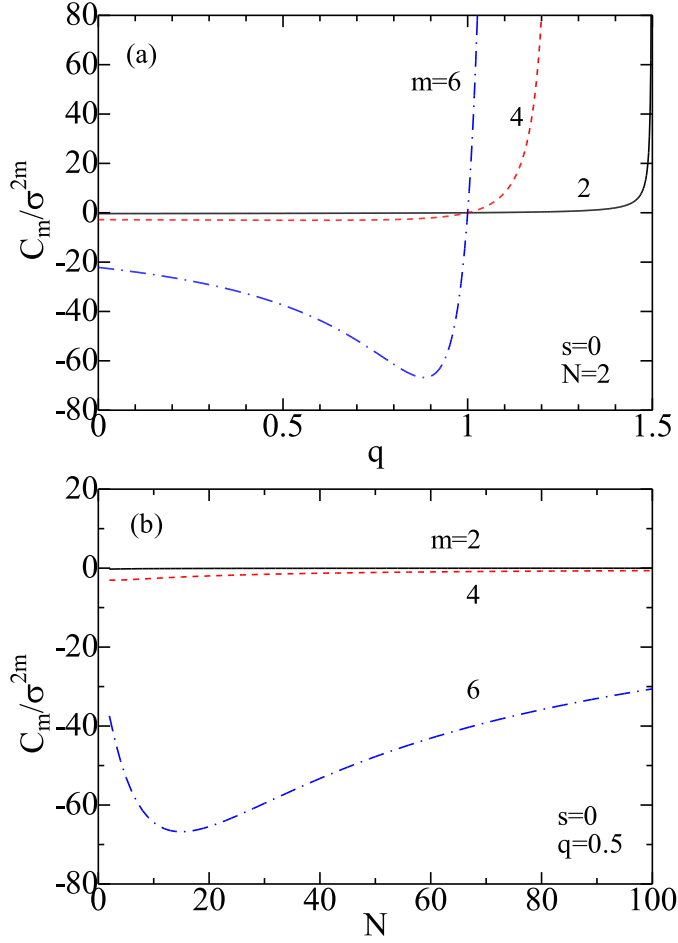}
\end{center}
\caption{
(Color online)
(a) The $q$ dependence of $C_m$ with $N=2$ and
(b) the $N$ dependence of $C_{m}$ with $q=0.5$ ($s=0$):
$m=2$ (the solid curve), 4 (the dashed curve) and 6 (the chain curve).
}
\label{fig4}
\end{figure}

Figure 3(a) shows the $q$ dependence of $C_{2s}$, $C_{2n}$ and $C_2$ 
($=C_{2s}+C_{2n}$) for $s=0.5$ and $N=2$.
We note that $C_{2s} \geq 0$, while $C_{2n} < 0$ for $q < 1.0$ and 
$C_{2n} \geq 0$ for $q \geq 1.0$.
The $q$ dependences of $C_2$ for $s= 0.0$ with $N=2$, 4, 10 and 20 are plotted
in Fig.3(b), where $C_2$ vanishes for $q=1.0$, and it is $-0.33$, $-0.25$, 
$-0.14$ and $-0.083$ for $N=2$, 4, 10 and 20, respectively, at $q=0.0$.
Figure 4(a) and 4(b) show the $q$- and $N$-dependent $C_m$,
respectively, for $m=2$, 4 and 6 with $s=0.0$.
Magnitudes of $C_m$ are significantly increased with increasing $m$.

\section{Discussion}

\subsection{PDF in the superstatistics}
The physical origin of the correlation induced by nonextensivity is easily 
understood in the superstatistics.
We consider the $N$-unit Langevin model subjected to additive noise 
given by \cite{Hasegawa08b}
\begin{eqnarray}
\frac{dx_i}{dt} &=& -\lambda x_i + \sqrt{2D} \:\xi_i(t)+I
\hspace{1cm}\mbox{for $i=1$ to $N$}, 
\end{eqnarray}
where $ \lambda $ denotes the relaxation rate,
$\xi_i(t)$ the white Gaussian noise with the intensity $D$,
and $I$ an external input.
The PDF of $\pi^{(N)}(\vecx)$ for the system is given by
\begin{eqnarray}
\pi^{(N)}(\vecx) &=& \prod_{i=1}^N \: \pi^{(1)}(x_i),
\label{eq:G0}
\end{eqnarray}
where the univariate PDF of $\pi^{(1)}(x_i)$ obeys the Fokker-Planck equation,
\begin{eqnarray}
\frac{\partial \pi^{(1)}(x_i,t)}{\partial t} 
&=& \frac{\partial }{\partial x_i}[(\lambda x_i -I) \pi^{(1)}(x_i,t)]
+ D \frac{\partial^2 }{\partial x_i^2} \pi^{(1)}(x_i,t).
\end{eqnarray}
The stationary PDF of $\pi^{(1)}(x_i)$ is given by
\begin{eqnarray}
\pi^{(1)}(x_i) &=& \frac{1}{\sqrt{2 \pi \sigma^2}}
\exp\left(-\frac{(x_i- \mu)^2}{2 \sigma^2} \right),
\label{eq:G1}
\end{eqnarray}
with 
\begin{eqnarray}
\mu=I/\lambda, \;\; \sigma^2=D/\lambda.
\end{eqnarray}

After the concept in the superstatistics \cite{Wilk00,Beck01,Beck05,Beck07},
we assume that a model parameter of $\tilde{\beta}$ ($\equiv \lambda/D$)
fluctuates, and that its distribution is expressed by 
the $\chi^2$-distribution with rank $n$ \cite{Wilk00,Beck01}, 
\begin{eqnarray}
f(\tilde{\beta}) &=& 
\frac{1}{\Gamma(n/2)}\left(\frac{n}{2\beta_0} \right)^{n/2}
\tilde{\beta}^{n/2-1} e^{-n \tilde{\beta}/2 \beta_0},
\end{eqnarray}
where $\Gamma(x)$ is the gamma function. Average and variance of $\tilde{\beta}$ 
are given by $ \langle \tilde{\beta} \rangle_{\tilde{\beta}}=\beta_0 $
and $(\langle \tilde{\beta}^2 \rangle_{\tilde{\beta}}-\beta_0^2)/\beta_0^2=2/n$, 
respectively.
Taking the average of $\pi^{(N)}(\vecx)$ over $f(\tilde{\beta})$,
we obtain the stationary PDF given by \cite{Hasegawa09}
\begin{eqnarray}
p_q^{(N)}(\vecx)&=& \int_0^{\infty} \pi^{(N)}(\vecx) \:f(\tilde{\beta})\:d\tilde{\beta},\\  
& = & \frac{1}{Z_q^{(N)}}
\exp_{q} \left[- \left( \frac{\beta_0}{2 \nu_q^{(N)}} \right) 
\sum_{i=1}^N (x_i-\mu)^2 \right],
\label{eq:G2}
\end{eqnarray}
with
\begin{eqnarray}
Z_q^{(N)} &=& 
\:\left[ \frac{2 \nu_q^{(N)}}{(q-1) \beta_0}\right]^{N/2}
\;\prod_{i=1}^N B\left(\frac{1}{2}, \frac{1}{q-1}-\frac{i}{2} \right), 
\label{eq:G3} 
\\
q &=& 1+ \frac{2}{(N+n)}, 
\label{eq:G5} 
\end{eqnarray}
where $\nu_q^{(N)}$ is given by Eq. (\ref{eq:C5}).
In the limit of $n \rightarrow \infty$ ($q \rightarrow 1.0$) where 
$f(\tilde{\beta}) \rightarrow \delta(\tilde{\beta}-\beta_0)$, the PDF reduces to 
the multivariate Gaussian distribution given by
\begin{eqnarray}
p_q^{(N)}(\vecx)  
&= & \frac{1}{(Z_1^{(1)})^N} \prod_{i=1}^N \exp \left[- \frac{\beta_0}{2}\: 
(x_i -\mu)^2 \right],
\label{eq:G6}
\end{eqnarray}
which agrees with Eqs. (\ref{eq:G0}) and (\ref{eq:G1}) 
for $\beta_0=\lambda/D=1/\sigma^2$.

We note that the PDF given by Eq. (\ref{eq:G2}) is equivalent to that
given by Eq. (\ref{eq:C6}) derived by the MEM when we read $\beta_0=1/ \sigma^2$.
The nonextensivility-induced correlation arises from 
the {\it common} fluctuating field of $\tilde{\beta}$, 
because $p_q^{(N)}(\vecx) \neq \prod_i p_q^{(1)}(x_i)$ for $q \neq 1.0$
despite $\pi^{(N)}(\vecx) = \prod_i \pi^{(1)}(x_i)$. 

\subsection{PDF for classical ideal gas}

It is worthwhile to discuss the PDF of one-dimensional ideal gas, whose
hamiltonian is given by
\begin{eqnarray}
H = \sum_{i=1}^N\: h_i = \sum_{i=1}^N\: \frac{p_i^2}{2m},
\end{eqnarray}
$m$ and $p_i$ standing for the mass and momentum, respectively, 
of ideal gas. 
By employing the OLM-MEM \cite{Martinez00}, we obtain the PDF given by
(for details, see Appendix B)
\begin{eqnarray}
p_q^{(N)}(\vecp) &=& \frac{1}{Z_q^{(N)}} 
\exp_q\left[- \left(\frac{\beta}{\nu_q^{(N)}}\right) 
\sum_{i=1}^N \:\frac{p_i^2}{2m} \right], 
\label{eq:H7}
\end{eqnarray}
with
\begin{eqnarray}
Z_q^{(N)} = \left\{ \begin{array}{ll}
\left[\frac{2 \pi\: m \:\nu_q^{(N)}}{(q-1) \beta} \right]^{N/2}
\frac{\Gamma\left(\frac{1}{q-1}-\frac{N}{2} \right)}
{\Gamma \left(\frac{1}{q-1} \right)}
\quad & \mbox{for $1 < q <3$}, \\
\left( \frac{2 \pi\: m}{\beta} \right)^{N/2}
\quad & \mbox{for $q=1$}, \\
\left[\frac{2 \pi m \:\nu_q^{(N)}}{(1-q) \beta} \right]^{N/2}
\frac{\Gamma\left(\frac{1}{1-q}+1 \right)}
{\Gamma \left(\frac{1}{1-q}+1+\frac{N}{2} \right)}
\quad & \mbox{for $ q <1$}. \\
\end{array} \right.
\label{eq:H8}
\end{eqnarray}
The internal energy is given by $U=N/2 \beta$ [Eq. (\ref{eq:H5})] 
independently of $q$, which
yields the Dulong-Petit specific heat as in the Boltzmann-Gibbs statistics.
The PDF of nonextensive ideal gas was originally discussed in Ref. \cite{Abe99}
with the use of the normalized MEM \cite{Tsallis98}. Later it was 
re-examined by the OLM-MEM  \cite{Martinez00}\cite{Abe01}.
The PDF given by Eq. (\ref{eq:H7}) is equivalent with the $q$-Gaussian PDF
given by Eq. (\ref{eq:C6}), if we read $m/\beta \rightarrow \sigma^2$.

\subsection{PDF for classical harmonic oscillator}

Next we consider the one-dimensional $N$-unit harmonic oscillators whose
hamiltonian is given by
\begin{eqnarray}
H = \sum_{i=1}^N\: h_i 
= \sum_{i=1}^N\: \left( \frac{p_i^2}{2m}+\frac{m \omega^2 x_i^2}{2} \right),
\end{eqnarray}
$m$, $\omega$, $x_i$ and $p_i$ expressing the mass, oscillator frequency, 
position and momentum, respectively. 
With the use of the OLM-MEM \cite{Martinez00}, the PDF is given by
(for details, see Appendix C)
\begin{eqnarray}
p_q^{(N)}(\vecp, \vecx) &=& \frac{1}{Z_q^{(N)}} 
\exp_q\left[- \left(\frac{\beta}{\nu_q^{(2 N)}}\right) \sum_{i=1}^N 
\:\left(\frac{p_i^2}{2m}+\frac{m \omega^2 x_i^2}{2} \right) \right],
\label{eq:K7}
\end{eqnarray}
with
\begin{eqnarray}
Z_q^{(N)} &=& \left\{ \begin{array}{ll}
\left[\frac{2 \pi \:\nu_q^{(2N)}}{(q-1) \omega \beta} \right]^{N}
\frac{\Gamma\left(\frac{1}{q-1}-N \right)}
{\Gamma \left(\frac{1}{q-1} \right)}
\quad & \mbox{for $1 < q <3$}, \\
\left( \frac{2 \pi}{\omega \beta} \right)^{N}
\quad & \mbox{for $q=1$}, \\
\left[\frac{2 \pi \:\nu_q^{(2N)} }{(1-q) \omega \beta} \right]^{N}
\frac{\Gamma\left(\frac{1}{1-q}+1 \right)}
{\Gamma \left(\frac{1}{1-q}+1+N \right)}
\quad & \mbox{for $ q <1$}, \\
\end{array} \right. 
\label{eq:K8}
\end{eqnarray}
where $\nu_q^{(2N)}=(N+1)-Nq$ [Eq. (\ref{eq:C5})].
The internal energy is given by $U=N/\beta$ [Eq. (\ref{eq:K5})], which
is the same as that in \cite{Liyan08} and in the Boltzmann-Gibbs statistics. 
The PDF of the harmonic oscillator was calculated
by using un-normalized \cite{Curado91} and normalized \cite{Tsallis98} MEMs
in Ref.\cite{Lenzi01}, where $U$ has rather complicated $q$ and $N$ dependences 
(see Eqs. (12) and (22) of Ref. \cite{Lenzi01}).

PDFs for classical ideal gas [Eqs. (\ref{eq:H7})] and harmonic oscillator 
[Eq. (\ref{eq:K7})] have the same structure as the $q$-Gaussian PDF given 
by Eq. (\ref{eq:C6}), and they are expected to have the same properties 
as the $q$-Gaussian PDF. Actually correlation defined by 
$\langle h_i h_j \rangle_q -\langle h_i \rangle_q \langle h_j \rangle_q$
($i \neq j$) is shown to be induced by nonextensivity
in ideal gas \cite{Abe99b,Liyan08,Feng10} and harmonic oscillator \cite{Feng10}.

\subsection{$q$-product PDF}

From functional forms of PDFs in Eq. (\ref{eq:H1}) or (\ref{eq:K1}), 
we expect that $p_q^{(N)}(\vecx)$ may be expressed by
\begin{eqnarray}
p_q^{(N)}(\vecx)  
&=& p_q^{(1)}(x_1) \otimes_q p_q^{(1)}(x_2) 
\otimes_q \cdot\cdot \otimes_q \: p_q^{(1)}(x_N),
\label{eq:L1}
\end{eqnarray}
where the $q$-product ($\otimes_q$) is defined by \cite{Borges04}
\begin{eqnarray}
x \otimes_q y &=& [x^{1-q}+y^{1-q}-1]^{1/(1-q)}.
\label{eq:L2}
\end{eqnarray}
Equation (\ref{eq:L1}), however, does not hold when we take into account the
precise form of PDFs including their normalization factors. 
For example, for a univariate PDF given by
\begin{eqnarray}
p_q^{(1)}(x) &=& \frac{1}{X_q^{(1)}} [1-(1-q) \beta h_1]^{1/(1-q)}, 
\label{eq:L3}
\end{eqnarray}
with
\begin{eqnarray}
X_q^{(1)} &=& {\rm Tr}_1 \;[1-(1-q) \beta h_1]^{1/(1-q)},
\end{eqnarray}
Eqs. (\ref{eq:L1}) and (\ref{eq:L3}) yield the $q$-product PDF for $N=2$,
\begin{eqnarray}
p_q^{(1)}(x_1) \otimes_q p_q^{(1)}(x_2)  
&= &\frac{1}{(X_q^{(1)})^{1-q}} 
\left[2-(X_q^{(1)})^{1-q}-(1-q)\beta (h_1+h_2) \right]^{1/(1-q)}.
\label{eq:L4}
\end{eqnarray}
Unfortunately, Eq. (\ref{eq:L4}) is not in agreement 
with the bivariate PDF given by
\begin{eqnarray}
p_q^{(2)}(x_1, x_2) 
&=& \frac{1}{X_q^{(2)}} [1-(1-q)\beta (h_1+h_2)]^{1/(1-q)}, 
\label{eq:L5}\\
&\neq & p_q^{(1)}(x_1) \otimes_q p_q^{(1)}(x_2), 
\end{eqnarray}
with
\begin{eqnarray}
X_q^{(2)} &=& {\rm Tr}_{12}\;[1-(1-q)\beta (h_1+h_2)]^{1/(1-q)}.
\label{eq:L6}
\end{eqnarray}
It is easy to see that the relation: 
$p_q^{(2)}(x_1,x_2)=p_q^{(1)}(x_1) \otimes_q p_q^{(1)}(x_2)$ holds only 
for $X_q^{(1)}=X_q^{(2)}=1$ in Eqs. (\ref{eq:L4}) and (\ref{eq:L5}).

\section{Concluding remarks}

It has been postulated that the Tsallis entropy $S_{q}^{(N)}$
satisfies the pseudoadditivity given by Eq. (\ref{eq:A2}) or (\ref{eq:D10}),
and that the factorized PDF:
$p_q^{(N)}(\vecx)=\prod_{i} p_q^{(1)}(x_i)$ leads to
the Tsallis entropy expressed by Eq. (\ref{eq:A1}) \cite{Tsallis01}.
The pseudoadditivity 
is a basis of the Tsallis entropy
and detailed discussions on its pseudoadditivity have been made
\cite{Tsallis01,Raj99,Abe00,Abe01b}.
We should note, however, that this is not self-consistent because
the PDF derived by the MEM for $S_{q}^{(N)}$ leads to
$p_q^{(N)}(\vecx) \neq \prod_{i} p_q^{(1)}(x_i)$
which contradicts with the postulated, factorized PDF.
The pseudoadditivity cannot be
a basis of the Tsallis entropy \cite{Lavenda05}.

It has been also controversial 
whether the $statistical$ $independence$ is expressed 
by (a) $p_q^{(N)}(\vecx)=\prod_{i} p_q^{(1)}(x_i)$
or (b) $H = \sum_{i} h_i$ 
in nonextensive systems \cite{Wang02,Wang02b,Jiulin09}. 
If we assume that the condition (a)
expresses the statistical independence of $N$-unit subsystems, 
we obtain $S_q^{(N)}=S_{q,FA}^{(N)}$ which satisfies the pseudoadditivity.
However, the PDF derived by the MEM for $S_q^{(N)}$
yields $p_q^{(N)}(\vecx) \neq \prod_{i} p_q^{(1)}(x_i)$, 
which is inconsistent with the assumption.
This means that
the statistical independence cannot be expressed by neither the product 
nor $q$-product PDFs.

Our calculations with the use of the multivariate $q$-Gaussian PDF 
have shown that

\noindent
(i) the Tsallis entropy is given by
$S_q^{(N)} =S_{q,FA}^{(N)}+\Delta S_q^{(N)} \geq S_{q,FA}^{(N)}$ where
the correction term of $\Delta S_q^{(N)}$
is significant for large $\vert q-1 \vert$ and/or large $N$, 

\noindent
(ii) the Tsallis entropy does not satisfy pseudoadditivity, and

\noindent 
(iii) nonextensivity-induced correlation is realized 
in higher-order correlations $C_m$ for $m \geq 2$ 
while $C_1$ expresses the intrinsic correlation.

\noindent
The items (i)-(iii) are expected to hold also for classical ideal gas
and harmonic oscillator.
The item (ii) is against the common wisdom \cite{Lavenda05}.
The nonextensivity-induced correlation in the item (iii)
is elucidated as arising from {\it common} fluctuating field introduced in
the superstatistics \cite{Wilk00,Beck01,Beck05}.
It has been shown that the FA is not a good approximating method 
in classical nonextensive systems, just as in quantum ones as recently 
pointed out in Refs. \cite{Hasegawa09b,Hasegawa09c}.
We should be careful in adopting the FA, although it has been widely 
employed in many applications of the classical and quantum nonextensive statistics
\cite{Nonext}.

\begin{acknowledgments}
This work is partly supported by
a Grant-in-Aid for Scientific Research from the Japanese 
Ministry of Education, Culture, Sports, Science and Technology.  
\end{acknowledgments}

\vspace{0.5cm}
\appendix*

\section{A. Evaluations of $q$-averages}
\renewcommand{\theequation}{A\arabic{equation}}
\setcounter{equation}{0}

We briefly discuss evaluations of $q$-averages given by
\begin{eqnarray}
Z_q^{(N)}(\alpha) &=&\int \: \left[1-(1-q)  
\alpha \:\Phi(\vecx) \right]^{1/(1-q)}\:d \vecx, \\
Q_q^{(N)}(\alpha) &=& \langle Q(\vecx) \rangle_q
=\frac{1}{\nu_q^{(N)} Z_q^{(N)}} \int \:Q(\vecx)
\: \left[1-(1-q) \alpha \:\Phi(\vecx) \right]^{q/(1-q)}\:d \vecx, 
\end{eqnarray}
by using the exact expressions for the gamma function 
\cite{Prato95,Rajagopal98,Hasegawa09b}:
\begin{eqnarray}
y^{-s} &=& \frac{1}{\Gamma(s)} \int_0^{\infty} u^{s-1}e^{-yu}\:du 
\hspace{2cm}\mbox{for $s > 0$}, \\
y^s &=&\frac{i}{2 \pi} \Gamma(s+1) \int_C (-t^{-s-1}) e^{-yt}\:dt
\hspace{1cm}\mbox{for $s > 0$},
\end{eqnarray}
where $\alpha=1/(2 \nu_q^{(N)} \sigma^2)$, $\Phi(\vecx)$ is given by
Eq. (\ref{eq:C0}), $Q(\vecx)$ denotes an arbitrary function of $\vecx$,
$C$ the Hankel path in the complex plane, 
and Eq. (\ref{eq:D2}) being employed.
We obtain \cite{Prato95,Rajagopal98,Hasegawa09b}
\renewcommand{\arraystretch}{2.0}
\begin{eqnarray}
Z_q^{(N)}(\alpha)
&=& \left\{ \begin{array}{ll}
\frac{1}{\Gamma\left[\frac{1}{q-1} \right]} 
\int_0^{\infty} u^{\frac{1}{q-1}-1} e^{-u}
Z_1^{(N)}[(q-1) \alpha u] \: du
\quad & \mbox{for $q > 1.0$}, \\
\frac{i}{2 \pi}\Gamma\left[\frac{1}{1-q}+1 \right] 
\int_C (-t)^{-\frac{1}{1-q}-1} e^{-t}
Z_1^{(N)}[-(1-q) \alpha t] \: dt
\quad & \mbox{for $q <  1.0$}, 
\end{array} \right. \nonumber \\
&& \label{eq:X1}
\end{eqnarray}
\begin{eqnarray}
Q_q^{(N)}(\alpha)
&=& \left\{ \begin{array}{ll}
\frac{1}{\nu_q^{(N)} Z_q^{(N)}
\Gamma\left[\frac{q}{q-1} \right]} \int_0^{\infty} u^{\frac{q}{q-1}-1} e^{-u}
Z_1^{(N)} [(q-1) \alpha u] \:Q_1^{(N)}[(q-1) \alpha u] \: du 
\hspace{1cm} & \mbox{for $q > 1.0$}, \\
\frac{i}{2 \pi \nu_q^{(N)} Z_q^{(N)}}
\Gamma\left[\frac{q}{1-q}+1 \right] \int_C (-t)^{- \frac{q}{1-q}-1} e^{-t}
Z_1^{(N)}[-(1-q) \alpha t] \\
\hspace{2cm} \times \:Q_1^{(N)}[-(1-q) \alpha t]\: dt 
\hspace{1cm} & \mbox{for $q <  1.0$}, 
\end{array} \right. \nonumber \\
&& \label{eq:X2}
\end{eqnarray}
where
\begin{eqnarray}
Z_1^{(N)}(\alpha) &=& \int e^{-\alpha \:\Phi(\vecx)}\:d\vecx, \\
%
Q_1^{(N)}(\alpha) &=& \frac{1}{Z_1^{(N)}(\alpha)}
\int \:Q(\vecx) \:e^{-\alpha \:\Phi(\vecx)}\:d\vecx.
\end{eqnarray}
Thus we may evaluate the $q$ average of $Q(\vecx)$ 
from its average over the Gaussian PDF.

For example, by using the relations for $N=2$, 
\begin{eqnarray}
\langle (x_i-\mu)^2 \rangle_1 &=& \frac{1}{2 \alpha}, \\
\langle (x_i-\mu) (x_j-\mu) \rangle_1
&=& \frac{s}{2 \alpha}
\hspace{2cm}\mbox{for $i \neq j$} \\
\langle (x_i-\mu)^2 (x_j-\mu)^2 \rangle_1
&=& \frac{1+ 2 s^2}{4 \alpha^2}
\hspace{1cm}\mbox{for $i \neq j$},
\end{eqnarray}
and employing Eqs. (\ref{eq:X1}) and (\ref{eq:X2}), we obtain
\begin{eqnarray}
\langle (x_i-\mu)^2 \rangle_q &=& \sigma^2, \\
\langle (x_i-\mu) (x_j-\mu) \rangle_q
&=& \sigma^2 s 
\hspace{5cm}\mbox{for $i \neq j$}, \\
\langle (x_i-\mu)^2 (x_j-\mu)^2 \rangle_q
&=& \frac{(N+2-Nq)(1+ 2 s^2) \: \sigma^4}
{(N+4)-(N+2)q}
\hspace{1cm}\mbox{for $i \neq j$},
\end{eqnarray}
which yield Eqs. (\ref{eq:E1})-(\ref{eq:E4}) and (\ref{eq:E5}).

\section{B. PDF for ideal gas}
\renewcommand{\theequation}{B\arabic{equation}}
\setcounter{equation}{0}

In order to obtain the PDF of $p_q^{(N)}(\vecp)$ for classical ideal gas
with the OLM-MEM \cite{Martinez00}, we impose the constraints given by
\begin{eqnarray}
1 &=& \int p_q^{(N)}(\vecp)\:d \vecp, 
\\
U &=& \sum_{i=1}^N \left< \frac{p_i^2}{2m} \right>_q.
\end{eqnarray}
The OLM-MEM \cite{Martinez00} yields
\begin{eqnarray}
p_q^{(N)}(\vecp) &\propto& \left[1-(1-q)\beta 
\left( \sum_{i=1}^N \:\frac{p_i^2}{2m} - U  \right) \right]^{1/(1-q)},
\label{eq:H1}
\end{eqnarray}
where the Lagrange multiplier $\beta$ expresses the inverse of temperature.
Rewriting Eq. (\ref{eq:H1}) as
\begin{eqnarray}
p_q^{(N)}(\vecp)
&\propto& \left[1-(1-q) 
\alpha \:\sum_{i=1}^N \:\frac{p_i^2}{2m} \right]^{1/(1-q)},
\label{eq:H2}
\end{eqnarray}
with
\begin{eqnarray}
\alpha &=& \frac{\beta}{1+(1-q)\beta U},
\label{eq:H3}
\end{eqnarray}
we obtain $U$ in terms of $\alpha$ as given by
\begin{eqnarray}
U &=& \frac{N}{(N+2-N q) \alpha }.
\label{eq:H4}
\end{eqnarray}
From Eqs. (\ref{eq:H3}) and (\ref{eq:H4}), $U$ and $\alpha$ are 
self-consistently determined as
\begin{eqnarray}
U &=& \frac{N}{2 \beta}, 
\label{eq:H5}\\
\alpha &=& \frac{2 \beta}{N+2-Nq}.
\label{eq:H6}
\end{eqnarray}
With the use of Eqs. (\ref{eq:H2}), (\ref{eq:H5}) and (\ref{eq:H6}), 
the PDF is given by Eqs. (\ref{eq:H7}) and (\ref{eq:H8}).

\section{C. PDF for harmonic oscillators}
\renewcommand{\theequation}{C\arabic{equation}}
\setcounter{equation}{0}

We obtain the PDF for classical harmonic oscillators
with the OLM-MEM \cite{Martinez00}, imposing the constraints given by
\begin{eqnarray}
1 &=& \int \int p_q^{(N)}(\vecp, \vecx)\:d \vecp\: d \vecx, 
\\
U &=& \sum_{i=1}^N \left< \frac{p_i^2}{2m} +\frac{m \omega^2 x_i^2}{2} \right>_q.
\end{eqnarray}
The OLM-MEM \cite{Martinez00} leads to
\begin{eqnarray}
p_q^{(N)}(\vecp, \vecx) &\propto& \left[1-(1-q)\beta 
\left( \sum_{i=1}^N \:\left( \frac{p_i^2}{2m} +\frac{m \omega^2 x_i^2}{2}\right)
- U  \right) \right]^{1/(1-q)},
\label{eq:K1}
\end{eqnarray}
which is rewritten as
\begin{eqnarray}
p_q^{(N)}(\vecp, \vecx)
&\propto& \left[1-(1-q) 
\alpha \:\sum_{i=1}^N 
\left( \:\frac{p_i^2}{2m}+\frac{m \omega^2 x_i^2}{2} \right) \right]^{1/(1-q)},
\label{eq:K2}
\end{eqnarray}
with
\begin{eqnarray}
\alpha &=& \frac{\beta}{1+(1-q)\beta U}.
\label{eq:K3}
\end{eqnarray}
We obtain $U$ in terms of $\alpha$ as given by
\begin{eqnarray}
U &=& \frac{N}{(N+1-N q) \alpha }.
\label{eq:K4}
\end{eqnarray}
From Eqs. (\ref{eq:K3}) and (\ref{eq:K4}), $U$ and $\alpha$ are 
self-consistently determined as
\begin{eqnarray}
U &=& \frac{N}{\beta}, 
\label{eq:K5}\\
\alpha &=& \frac{\beta}{N+1-Nq}.
\label{eq:K6}
\end{eqnarray}
By using Eqs. (\ref{eq:K2}), (\ref{eq:K5}) and (\ref{eq:K6}), 
we finally obtain the PDF given by
Eqs. (\ref{eq:K7}) and (\ref{eq:K8}).





\begin{thebibliography}{99}

\bibitem{Tsallis88}C. Tsallis:
J. Stat. Phys. {\bf 52}, 479 (1988).

\bibitem{Tsallis98}C. Tsallis, R. S. Mendes, 
and A. R. Plastino: Physica A {\bf 261}, 534 (1998). 

\bibitem{Tsallis01}C. Tsallis,
in {\it Nonextensive Statistical Mechanics and Its Application},
edited by S. Abe and Y. Okamoto (Springer-Verlag, Berlin, 2001), p 3.

\bibitem{Tsallis04}C. Tsallis:
Physica D {\bf 193}, 3 (2004).

\bibitem{Curado91}E. M. F. Curado and C. Tsallis,
J. Phys. A {\bf 24} (1991) L69; {\bf 24}, 3187 (1991);
{\bf 25}, 1019 (1992).


\bibitem{Martinez00}S. Martinez, F. Nicolas, F. Pennini,
and A. Plastino,
Physica A {\bf 286}, 489 (2000).

\bibitem{Ferri05}G. L. Ferri, S. Martinez, and A. Plastino,
J. Stat. Mech. Theory Exp., p04009 (2005).


\bibitem{Wang02}Q. A. Wang, M. Pezeril, L. Nivanen, 
and A. L. M\'{e}haut\'{e},
Chaos, Solitons and Fractals {\bf 13}, 131 (2002).

\bibitem{Wang02b}Q. A. Wang,
Physics Lettrs A {\bf 300}, 169 (2002).

\bibitem{Jiulin09}D. Jiulin,
arXiv:0906.1409


\bibitem{Lenzi01}E. K. Lenzi, R. S. Mendes, L. R. da Silva, and L. C. Malacarne,
Physica A {\bf 289}, 44 (2001).

%
\bibitem{Hasegawa08b}H. Hasegawa,
Phys. Rev. E {\bf 78}, 021141 (2008).


\bibitem{Hasegawa09}H. Hasegawa,
Phys. Rev. E {\bf 80}, 051125 (2009).

\bibitem{Abe99b}S. Abe, 
Physica A {\bf 269}, 403 (1999).

\bibitem{Liyan08}L. Liyan and D. Jiulin,
Physica A {\bf 387}, 5417 (2008).

\bibitem{Feng10}Zhi-Hui Feng and Li-Yan Liu,
Physica A {\bf 389}, 237 (2010).



\bibitem{Wilk00}G. Wilk and Z. Wlodarczyk,
Phys. Rev. Lett. {\bf 84}, 2770 (2000).

\bibitem{Beck01}C. Beck,
Phys. Rev. Lett. {\bf 87}, 180601 (2001).

\bibitem{Beck05}C. Beck and E. G. D. Cohen,
Physica A {\bf 322}, 267 (2003).

\bibitem{Beck07}C. Beck,
arXiv:0705.3832.


\bibitem{Borges04}E. P. Borges,
Physica A {\bf 340}, 95 (2004).


\bibitem{Prato95}D. Prato,
Phys. Lett. A {\bf 203}, 165 (1995).

\bibitem{Rajagopal98}A. K. Rajagopal, R. S. Mendes and E. K. Lenzi, 
Phys. Rev. Lett. {\bf 80}, 3907 (1998).

\bibitem{Abe02}S. Abe,
Phys. Rev. E {\bf 66}, 046134 (2002).


\bibitem{Abe99}S. Abe, 
Phys. Lett. A {\bf 263}, 424 (1999);  {\bf 267}, 456 (2000)[E].

\bibitem{Abe01}S. Abe, S. Martinez, F. Pennini, and A. Plastino,
Physics Letters A {\bf 278}, 249 (2001).

\bibitem{Raj99}A. K. Rajagopal and S. Abe,
Phys. Rev. Lett. {\bf 83}, 1711 (1999).

\bibitem{Abe00}S. Abe, 
Phys. Lett. A {\bf 271}, 74 (2000).

\bibitem{Abe01b}S. Abe, 
Phys. Rev. E {\bf 63}, 061105 (2001).

\bibitem{Lavenda05}B. H. Lavenda and J. Dunning-Davies,
J. Appl. Sci. {\bf 5}, 920 (2005).

\bibitem{Hasegawa09b}H. Hasegawa,
Phys. Rev. E {\bf 80}, 011126 (2009).

\bibitem{Hasegawa09c}H. Hasegawa,
arXiv:09060225.

\bibitem{Nonext}Lists of many applications of the nonextensive
statistics are available at 
URL: \\
(http://tsallis.cat.cbpf.br/biblio.htm)

\end{thebibliography}
\end{document}